\documentclass[twocolumn]{aastex631}
\usepackage{amsmath}

\newcommand{\Mjup}{$M_\mathrm{Jup}$}

\newcommand{\Rp}{$R_\mathrm{p}$}

\newcommand{\kms}{km\,s$^{-1}$}
\newcommand{\K}{KELT-9}

\newcommand{\caltech}{Department of Astronomy, California Institute of Technology, Pasadena, CA 91125, USA}
\newcommand{\gps}{Division of Geological \& Planetary Sciences, California Institute of Technology, Pasadena, CA 91125, USA}
\newcommand{\ipac}{IPAC, Mail Code 100-22, Caltech, 1200 E. California Boulevard, Pasadena, CA 91125, USA}
\newcommand{\carnegiew}{Earth and Planets Laboratory, Carnegie Institution for Science, Washington, DC, 20015}
\newcommand{\hawaii}{Institute for Astronomy, University of Hawaii at M\=anoa, 2680 Woodlawn Drive, Honolulu, HI 96822, USA}
\newcommand{\mitech}{Department of Earth, Atmospheric and Planetary Sciences, Massachusetts Institute of Technology, Cambridge, MA 02139, USA}
\newcommand{\mitkavli}{Kavli Institute for Astrophysics and Space Research, Massachusetts Institute of Technology, Cambridge, MA 02139, USA}
\newcommand{\berkeley}{Department of Astronomy, University of California Berkeley, Berkeley, CA 94720, USA}
\newcommand{\uclagps}{Department of Earth, Planetary, and Space Sciences, University of California, Los Angeles, CA 90095, USA}
\newcommand{\nortredame}{Department of Physics and Astronomy, University of Notre Dame, Notre Dame, IN, 46556, USA}
\newcommand{\ucsb}{Department of Physics, University of California, Santa Barbara, CA 93106, USA}
\newcommand{\shao}{Shanghai Astronomical Observatory, Chinese Academy of Sciences, Shanghai 200030, China}
\newcommand{\imperial}{Imperial Astrophysics, Imperial College London, South Kensington Campus, London SW7 2AZ, UK}
\newcommand{\chicago}{Department of Astronomy \& Astrophysics, University of Chicago, Chicago, IL 60637, USA}
\newcommand{\yale}{Department of Astronomy, Yale University, New Haven, CT 06511, USA}

\shorttitle{Hydrogen airglow}
\shortauthors{Zhang et al.}

\graphicspath{{./}{figures/}}

\begin{document}

\title{Hydrogen airglow from an escaping ultrahot Jupiter atmosphere}

\author[0000-0003-0097-4414]{Yapeng Zhang}
\altaffiliation{51 Pegasi b Fellow}
\affiliation{\caltech}

\author{Chenliang Huang}
\affiliation{\shao}

\author[0000-0002-5812-3236]{Aaron Householder}
\altaffiliation{NSF Graduate Research Fellow}
\affiliation{\mitech}
\affiliation{\mitkavli}

\author{James E. Owen}
\affiliation{\imperial}

\author{Fei Dai}
\affiliation{\hawaii}

\author[0000-0002-3239-5989]{Aurora Y. Kesseli}
\affiliation{\ipac}

\author[0000-0001-8638-0320]{Andrew W. Howard}
\affiliation{\caltech}

\author{Julie Inglis}
\affiliation{\gps}

\author[0000-0002-0531-1073]{Howard Isaacson}
\affiliation{\berkeley}

\author[0000-0002-5375-4725]{Heather A. Knutson}
\affiliation{\gps}

\author{Dimitri Mawet}
\affiliation{\caltech}

\author{Nicole Wallack}
\affiliation{\carnegiew}

\author[0000-0002-6618-1137]{Jerry W. Xuan}
\altaffiliation{51 Pegasi b Fellow}
\affiliation{\caltech}
\affiliation{\uclagps}

\author{Michael Zhang}
\affiliation{\chicago}

\author[0000-0001-6416-1274]{Theron W. Carmichael}
\altaffiliation{NSF Ascend Postdoctoral Fellow}
\affiliation{\hawaii}

\author{Daniel Huber}
\affiliation{\hawaii}

\author{Rena A. Lee}
\affiliation{\hawaii}

\author[0000-0003-2657-3889]{Nicholas Saunders}
\affiliation{\yale}
\affiliation{\hawaii}

\author{Lauren M. Weiss}
\affiliation{\nortredame}

\author{Jingwen Zhang}
\affiliation{\hawaii}
\affiliation{\ucsb}

\begin{abstract}

Intense high-energy irradiation of close-in gaseous exoplanets drives the rapid escape of their atmospheres, fundamentally shaping planetary demographics.
While atmospheric loss is routinely observed via transit absorption in atomic hydrogen, helium, and metal ions, the underlying physical properties, specifically the thermal structure, outflow dynamics, and mass-loss rate, remain poorly constrained due to inherent degeneracies in the transmission geometry.
Here we report the first detection of atomic hydrogen emission from the escaping atmosphere of a gas giant.
Using high-resolution spectroscopy of the ultrahot Jupiter \K~b, we detect a hydrogen Balmer line (H$\alpha$ 6564.6 \AA) emission signature originating from the planetary dayside.
The emission line profile features a distinctive double-peaked shape with $0.1$--$0.15\%$ peak amplitudes at $\pm 30$ \kms~and central self-absorption.
This profile breaks transmission degeneracies, providing direct observational constraints on the vertical thermal structure, excited-state hydrogen populations, and wind dynamics in the upper atmosphere of \K~b.
Initial modeling reveals a vigorous outflow with a mass-loss rate above $10^{13}$ g\,s$^{-1}$, among the highest measured to date for gaseous exoplanets.
Our results establish hydrogen airglow emission as a powerful diagnostic of atmospheric escape, opening a new observational window into the evolution of worlds in extreme radiation environments.

\end{abstract}


\section{Introduction} \label{sec:intro}

\K~b is the hottest known exoplanet, orbiting an A0 star every 1.48 days on a nearly polar orbit with an equilibrium temperature of $\sim$4,000~K \citep{GaudiEtAl2017, AhlersEtAl2020, WongEtAl2020, MansfieldEtAl2020}.
Transmission spectroscopy in the hydrogen Balmer and Paschen series has revealed an extended envelope susceptible to escape \citep{YanHenning2018, CauleyEtAl2019, WyttenbachEtAl2020, Sanchez-LopezEtAl2022}, and near-ultraviolet observations with the Hubble Space Telescope have found Mg\,{\sc ii} and Fe\,{\sc ii} absorption extending beyond the planet's Roche lobe, indicating ongoing mass loss \citep{BaldwinEtAl2025}.

While extreme-ultraviolet (EUV; $\lambda < 912$ \AA) radiation is believed to drive atmospheric escape in hot Jupiters orbiting late-type stars, Balmer-driven escape is proposed to dominate in UHJs \citep{FossatiEtAl2018, MunozSchneider2019}. This shift in mechanism stems from the distinct spectral energy distributions of early-type host stars, which exhibit relatively weak EUV but intense near-ultraviolet (NUV; $\lambda < 4000$ \AA) emission. Consequently, NUV radiation becomes the primary energy source deposited into the UHJ upper atmosphere and drives the escape \citep{MunozSchneider2019}. This process would lead to higher mass-loss rates than EUV-driven models, significantly impacting the long-term evolution of these atmospheres \citep{MunozSchneider2019, FossatiEtAl2020, WyttenbachEtAl2020, YanEtAl2021a, HuangEtAl2023}.

Transmission spectra, though rich in atomic tracers, suffer from an inherent degeneracy between thermospheric temperature, gas density, and mass-loss rate \citep{WyttenbachEtAl2020, LinssenEtAl2022}.
Emission from the escaping gas could break this degeneracy by probing the vertical temperature and density structure directly, but such emission has been predicted to be faint and previous searches in other hot Jupiters were inconclusive \citep{MenagerEtAl2013, RosenerEtAl2025, ZhangEtAl2020}.

Here we present the detection of H$\alpha$ (6564.6 \AA) emission from the dayside of \K~b using the high-resolution optical spectrograph Keck Planet Finder \citep[KPF;][]{GibsonEtAl2020a, GibsonEtAl2024} on the Keck I telescope.
Such emission is the exoplanet analogue of airglow, the faint optical glow produced in Earth's upper atmosphere as atoms relax from excited states populated by solar radiation.
Our observations spectrally resolved the line into a distinctive double-peaked profile with central self-absorption, which records the thermal structure, the excited-state hydrogen population, and the wind dynamics of the escaping atmosphere.
We interpret this profile with a one-dimensional non-local-thermodynamic-equilibrium (NLTE) outflow model and use it to place the first emission-based constraint on the escaping atmosphere of a hot Jupiter.

\section{Observations and Data Reduction} \label{sec:observation}

We observed emission from \K~b across multiple orbital phases using KPF on UT 2024 August 8, 20, and 27 (PI: Y. Zhang).
KPF is a high-resolution echelle spectrometer mounted on the Keck I telescope covering the optical wavelength range 445--870 nm with a resolving power of $\mathcal{R}\sim 98\,000$.
Given the planet's large orbital semi-amplitude ($K_\mathrm{p}\sim 245$ \kms), we used an exposure time of 150 s to minimize phase smearing of the planetary signal.
Observing conditions were favourable, with airmass between 1.07 and 1.70, and average seeing ranging from 0.3\arcsec~on August 8 to 0.5\arcsec~on August 20 and 27 in $V$-band.
This yielded a typical signal-to-noise ratio (S/N) per pixel of $\sim$400, 330, and 320 at 650 nm for the three nights, respectively.
Total on-target integration times were 3.5~h (85 exposures), 2.5~h (59 exposures), and 5.2~h (125 exposures).
The observations span a range of orbital phases: $\phi=0.30$--$0.42$ (Night 1), $\phi=0.40$--$0.48$ (Night 2), and $\phi=0.05$--$0.25$ (Night 3); see Fig.~\ref{fig:app1}.

We used spectra processed by the KPF Data Reduction Pipeline \citep{GibsonEtAl2020a}, which performs standard flat-fielding, order tracing, and optimal extraction.
We further performed blaze correction using the blaze functions extracted from the smooth lamp pattern \citep{HouseholderEtAl2025}.
Wavelength calibration was achieved using laser-frequency-comb and thorium-argon exposures bracketing the science observations.
Telluric absorption features from H$_2$O and O$_2$ were corrected using the ESO sky tool \texttt{Molecfit} \citep{SmetteEtAl2015}.
Figure~\ref{fig:app1} demonstrates the telluric removal for the two spectral orders covering 652--662 nm.

The telluric-corrected spectra were shifted to the stellar rest frame by accounting for the systemic and barycentric velocities and the star's reflex motion.
A master stellar spectrum was generated for each night by average-combining the respective time series.
Individual exposures were divided by the nightly master spectrum to remove stationary stellar features.
Observational uncertainties were estimated from the standard deviation of the residuals in each wavelength channel.

We focus on the two spectral orders that each independently cover the H$\alpha$ line.
Around the stellar H$\alpha$ line centre, we identified a prominent residual pattern caused by stellar pulsation, as previously reported in photometric and spectroscopic studies \citep{WongEtAl2020, WyttenbachEtAl2020}.
This pulsation signal contaminates the data within the stellar rotational velocity profile ($v\sin i\sim 110$ \kms).
While the planetary radial velocity exceeds this threshold at most orbital phases, the stellar contamination overlaps with the planetary track at $\phi>0.4$.
We therefore masked the data within the affected velocity range ($|v|<110$ \kms~in the stellar rest frame) to guarantee an unbiased extraction of the planetary signature.

\section{Detection of H$\alpha$ emission} \label{sec:detection}

As shown in Fig.~\ref{fig:app1}, the two-dimensional time-series residuals after telluric and stellar removal show a pair of bright, curved traces that shift with the expected Keplerian motion of the planet.
This Doppler signature unambiguously identifies the excess emission as planetary in origin.
Because the wavelength region around H$\alpha$ is simultaneously covered by two independent spectral orders of KPF, we can confidently rule out instrumental artefacts as we detect consistent features in both orders.

\begin{figure}[t]
    \centering
    \includegraphics[width=\linewidth]{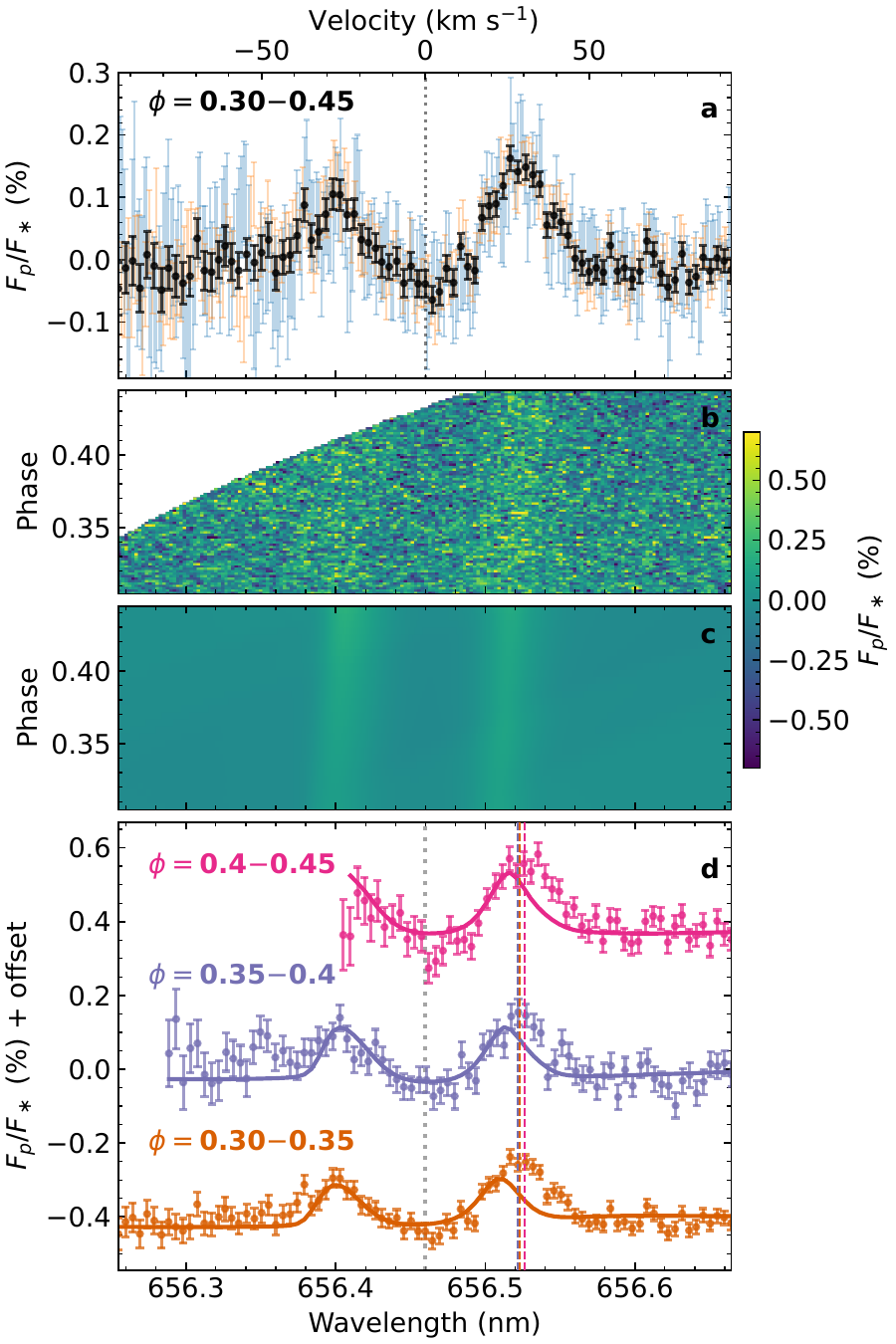}
    \caption{H$\alpha$ emission line in \K~b.
    \textbf{a}, Phase-averaged H$\alpha$ emission over $\phi=0.30$--$0.45$. The double-peaked emission with central self-absorption is detected at $10\sigma$. Blue and orange points show the two independent spectral orders; black points are binned fluxes in every five data points.
    \textbf{b}, 2D time-series H$\alpha$ spectra in the planet's rest frame; the two emission peaks appear in slanted bright yellow traces.
    Wavelengths contaminated by stellar pulsation are masked, which led to the data truncation on the top left. 
    \textbf{c}, Synthetic 2D time-series model spectra. The peak locations shift redward by $\sim$3.8 \kms~from the bottom to the top.
    \textbf{d}, Observed and modeled spectra averaged in three phase bins ($0.30$--$0.35$, $0.35$--$0.40$, $0.40$--$0.45$). The gray dotted line is the central wavelength of H$\alpha$. The colored vertical dashed lines annotate the red peak’s location for each phase bin. 
    The model reproduces the progressive redshift of the peaks with phase, driven by planetary rotation and day-to-night winds.
    \label{fig:ha_main}}
\end{figure}

Shifting the spectra into the planet's rest frame using a semi-amplitude of $K_\mathrm{p}=244.8$ \kms~\citep{ZhangEtAl2026}, and averaging across phases where the dayside dominates the visible hemisphere ($\phi=0.30$--$0.45$), we detect the integrated H$\alpha$ emission at $10\sigma$ significance, with an equivalent width of $0.00080\pm0.00008$ \AA, as shown in Fig.~\ref{fig:ha_main}a.
The combined line profile is intriguingly non-Gaussian: it shows two emission peaks at $\pm 30$ \kms, with the line core absorbed down to the continuum.
The peaks are asymmetric, with a stronger red wing (amplitude $\sim 0.15\%$) and a weaker blue wing ($\sim 0.10\%$).
No significant signal is recovered at phases where most of the dayside is hidden ($\phi=0.10$--$0.25$, Fig.~\ref{fig:ha_1d}), as expected if the emission originates from the irradiated hemisphere.

Subdividing the detection into finer phase bins reveals the phase-dependence of the emission signal.
The emission feature shifts progressively redward as the orbital phase advances towards secondary eclipse (Fig.~\ref{fig:ha_main}b\&c).
The extent of the shift is $\sim$3.8 \kms~from $\phi\sim0.3$ to 0.45 in the 2D model.
The same trend has been observed in dayside metal lines of \K~b, which is attributed to the planet rotation and horizontal winds \citep{ZhangEtAl2026}; see also Section~\ref{sec:origin}.

\section{NLTE outflow model}\label{sec:model}

\subsection{Outflow structure}

To interpret the double-peaked morphology, we developed a one-dimensional NLTE outflow model of \K~b. The model uses a customized version of the \texttt{sunbather} code \citep{LinssenEtAl2024}, with an iterative coupling between the radiative energy balance solved by the NLTE \texttt{Cloudy} code \citep[c23.01,][]{GunasekeraEtAl2023} and the mass and momentum continuity equations. 

The original \texttt{sunbather} adopts an isothermal Parker wind density and velocity profile assuming a single temperature $T_\mathrm{iso}$ and mass-loss rate $\dot{M}$, generated via the \texttt{p-winds} package \citep{DosSantosEtAl2022}.
\texttt{Cloudy} then solves the energy balance and computes a non-isothermal temperature and ionization structure, accounting for radiative heating and cooling from atomic species given the input stellar irradiation.
However, in the default \texttt{sunbather} framework, the underlying isothermal Parker wind structure is not self-consistent with the non-isothermal temperature profile returned by \texttt{Cloudy}.
Because H$\alpha$ emission is sensitive to both thermal structure and density profile, this inconsistency can bias the predicted line shape.
We therefore introduced an iterative scheme to achieve self-consistency between the outflow structure and the radiative energy balance.

Starting from an isothermal Parker wind profile, we run \texttt{Cloudy} to obtain a converged temperature structure $T(r)$ and mean molecular weight $\mu(r)$. We then update the density $\rho(r)$ and velocity $v(r)$ profiles by solving the steady-state mass and momentum continuity equations:
\begin{equation}
    \frac{\partial}{\partial r}\left(r^2 \rho v\right)=0,
\end{equation}
\begin{equation}
    v \frac{\partial v}{\partial r}=-\frac{1}{\rho}\frac{\partial P}{\partial r}-\frac{G M_{\mathrm{p}}}{r^2}+\frac{3 G M_* r}{a^3},
\end{equation}
with the boundary condition set by the input mass-loss rate,
\begin{equation}
    \dot{M} = 4\pi r^2 \rho v.
\end{equation}
The updated outflow structure is fed back into \texttt{Cloudy} and the cycle is repeated until convergence, typically reached within five iterations.
The converged profiles are insensitive to the initial choice of $T_\mathrm{iso}$, effectively eliminating it as a free parameter and leaving $\dot{M}$ as the sole input controlling the outflow structure.

We ran this framework for \K~b across a grid of $\dot{M}$ from $10^{11}$ to $10^{14}$ g\,s$^{-1}$ in 0.5-dex increments.
Atmospheric profiles are computed from 1 to 3 \Rp, beyond which hydrogen is mostly ionized and contributes negligibly to the emission.
We adopt a PHOENIX model with $T_\mathrm{eff}=10\,000$ K as the input stellar spectrum \citep{HusserEtAl2013}, since A-type stars are expected to have little extreme-ultraviolet (XUV) emission due to the lack of chromosphere/corona \citep{FossatiEtAl2018}, and \K~only has an reported upper limit on the XUV luminosity \citep{Sanz-ForcadaEtAl2025}.
Instead, the NUV flux has been proposed as the dominant energy source driving the outflow for planets orbiting A-type stars \citep{MunozSchneider2019}.
This NUV energy budget is incorporated in \texttt{Cloudy} to compute the outflow structure self-consistently.

\texttt{Cloudy} calculations become unreliable at number densities above $10^{15}$ cm$^{-3}$ \citep{FossatiEtAl2021}, which is the case in the lower atmosphere of \K~b.
We therefore mask the high-pressure levels ($P>10^{-3}$ bar) in high-$\dot{M}$ cases, and extrapolate the resulting temperature profile to the masked lower atmosphere based on literature studies of the hydrostatic atmosphere, which suggest a dayside photospheric temperature of $\sim$4,500~K and a steep temperature gradient \citep{ZhangEtAl2026, FossatiEtAl2021}.
Including the lower atmosphere is essential for determining the continuum level of the H$\alpha$ emission.

\begin{figure*}[t]
    \centering
    \includegraphics[width=0.9\linewidth]{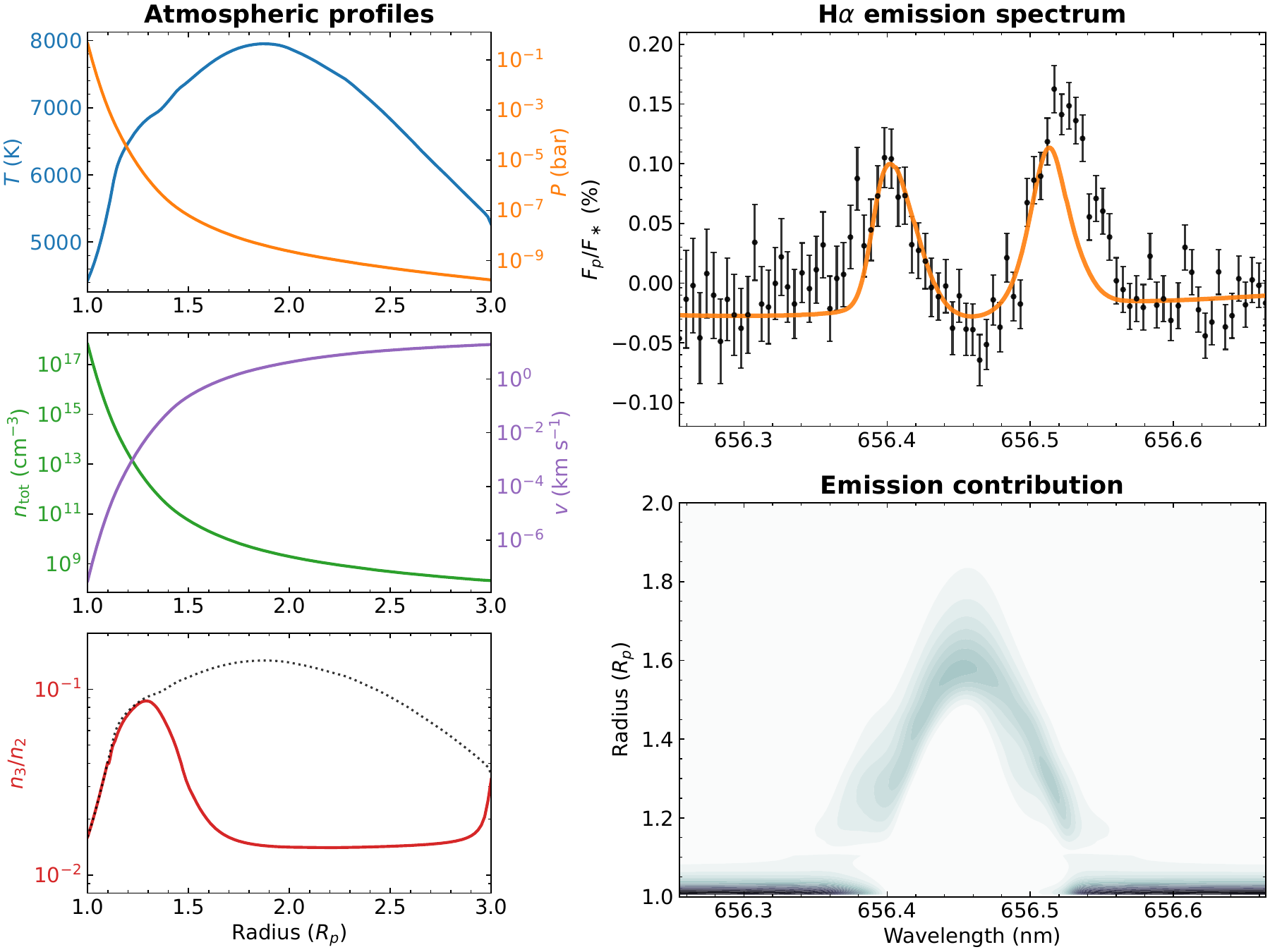}
    \caption{Left column shows the outflow temperature, pressure, velocity, and hydrogen number density profiles as a function of radius for $\dot{M}=10^{13}$ g\,s$^{-1}$, together with the excited-state population ratio $n_3/n_2$. The dotted black line shows $n_3/n_2$ under the LTE assumption, highlighting the NLTE departure in the extended atmosphere.
    Right column shows the outflow model synthetic spectrum for $\dot{M}=10^{13}$ g\,s$^{-1}$ (orange) compared with the observed line profile (black) and the emission contribution function as a function of wavelength and radius. The line core probes up to 1.8 \Rp, while the double peaks probe $\sim$1.3 \Rp.
    \label{fig:best_fit}}
\end{figure*}

\subsection{Synthetic emission spectra}

We then computed disk-integrated synthetic emission spectra accounting for the gas dynamics induced by planetary rotation, radial outflow, and day-to-night zonal winds.

Using the \texttt{Cloudy}-produced temperature structure $T(r)$ and the state-specific H\,{\sc i} number density profiles $n_2(r)$ and $n_3(r)$, we compute the synthetic spectrum following a plane-parallel radiative transfer scheme:
\begin{equation} \label{eq:intensity}
    I = S_0 e^{-\tau_0} + \int_0^{\tau_0} e^{-\tau} S \, d\tau,
\end{equation}
where $\tau$ is the optical depth and $S$ is the source function,
\begin{equation} \label{eq:source}
    S = \frac{\epsilon_\nu}{\kappa_\nu} = 2hc^2\nu_0^3 \frac{g_i n_j}{g_j n_i - g_i n_j},
\end{equation}
with $i$ and $j$ labelling the $n=2$ and $n=3$ excited states of H\,{\sc i}. $\tau_0$ and $S_0$ are the optical depth and source function at the lower boundary at 1 \Rp.

In addition to the H\,{\sc i} Balmer transition, the H$^-$ bound-free transition is the dominant source of continuous opacity at visible wavelengths.
We include H$^-$ opacity in our optical-depth calculation with the bound-free cross-sections of \cite{Gray2005}; the H$^-$ volume mixing ratio is computed with \texttt{easychem} \citep{LeiMolliere2024} given the pressure--temperature profile.

To obtain the disk-integrated emergent spectrum, we discretize the planetary surface into a $48\times 24$ longitude--latitude ($\varphi$, $\theta$) grid. For a given orbital phase $\phi$, we evaluate the emission intensity from individual grid cells along the radial direction, scaling the optical depth by the viewing geometry $\tau/\mu_\Theta$, with
\begin{equation}
    \mu_\Theta = \mathbf{\hat{n}} \cdot \mathbf{\hat{k}} = -\cos\theta \cos(\phi+\varphi).
\end{equation}
The total emission is the sum of the intensity over all patches weighted by their projected areas.
We assume uniform intensity on the dayside and zero contribution from the nightside.

Moreover, we account for local Doppler shifts induced by planetary rotation and atmospheric winds.
We assume the rotational velocity of the extended atmospheric layers scales as $1/r$ to conserve angular momentum, anchored to a 1.48-day solid-body rotation period at 1 \Rp.
Day-to-night zonal winds are parameterized as $v_\varphi=u_\varphi\sin\varphi$, with $u_\varphi=11.6$ \kms~adopted from metal-line measurements in \K~b's lower atmosphere \citep{ZhangEtAl2026}.
Isotropic Parker winds along the radial direction are also included in the line-of-sight velocity calculation.

We compute emergent synthetic spectra at three discrete phases ($\phi=0.30$, $0.375$, and $0.45$) and interpolate across the full phase range.
The synthetic spectra are then subjected to the same data reduction steps as the observations, including continuum normalization, masking of stellar-contaminated wavelength channels, and phase-averaging. 
This step replicates any potential self-subtraction of the planetary signal due to the data reduction procedure, so that it did not bias our model interpretation. 
All model spectra are computed over a $\pm 100$ \kms~velocity window centred on the H$\alpha$ line and convolved to the KPF instrumental resolution ($\mathcal{R}\sim 98\,000$).

\section{Origin of the double-peaked profile} \label{sec:origin}

A model with a mass-loss rate $\dot{M}=10^{13}$ g\,s$^{-1}$ or higher provides a broadly good match to the observed line profile (Fig.~\ref{fig:model_grid}), reproducing both the amplitude and the double-peaked morphology. The corresponding atmospheric profiles (Fig.~\ref{fig:best_fit}) show a temperature inversion rising from $\sim$4,500~K at the photosphere to $\sim$8,000~K near 1.9 \Rp, followed by expansion cooling of the escaping gas, and an NLTE departure of the excited-state hydrogen populations above $\sim$1.3 \Rp. Together these features produce a unique vertical structure in the H$\alpha$ source function, which in turn drives the observed line shape through the following physical effects (Fig.~\ref{fig:schematic}):

\begin{figure*}[t]
    \centering
    \includegraphics[width=0.7\linewidth]{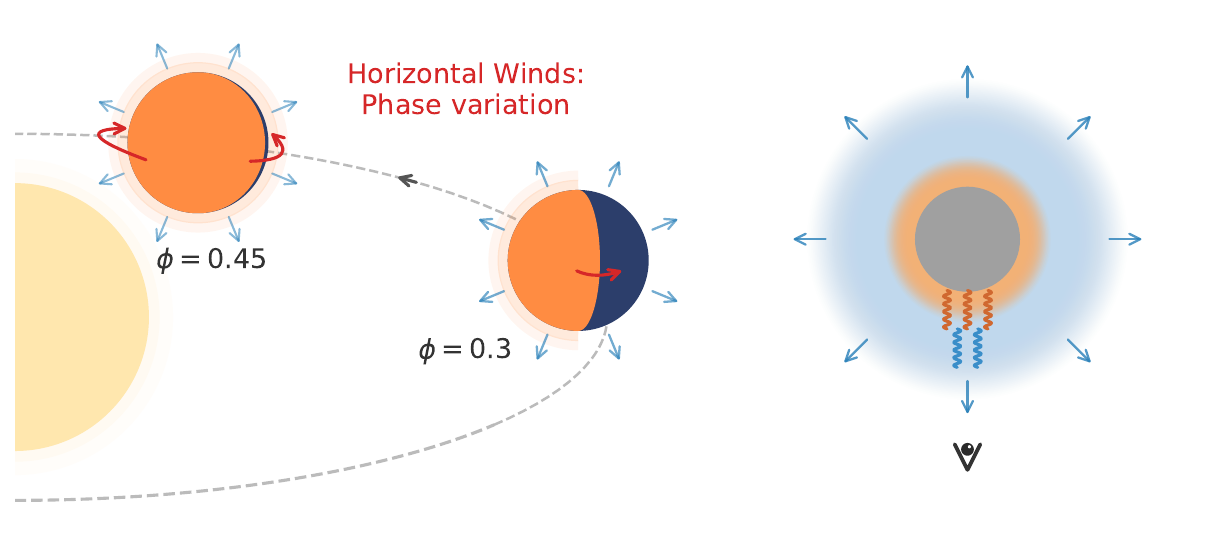}
    \includegraphics[width=0.7\linewidth]{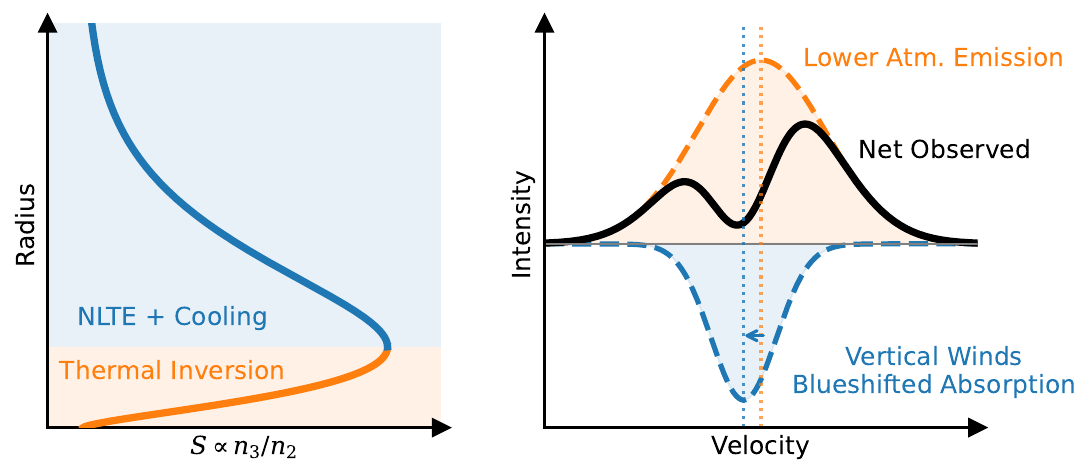}
    \caption{Physical origin of the double-peaked H$\alpha$ emission. The hot ($>$4,500~K) lower atmosphere shows a thermal inversion and emits in H$\alpha$ (orange). The extended upper atmosphere has a decreasing source function ($n_3/n_2$) and a temperature fall-off due to NLTE effects and expansion cooling, and absorbs the line core (blue). The vertical outflow wind blueshifts the absorbing layer relative to the emitting layer, producing asymmetric double peaks. Horizontal day-to-night winds advect gas away from the observer as the dayside hemisphere rotates into view, producing the progressive redshift with orbital phase.
    \label{fig:schematic}}
\end{figure*}

\begin{itemize}

\item \textit{Temperature inversion drives emission.} If Lyman~$\alpha$ photons produced by thermal excitation in the lower atmosphere cannot escape due to collisional de-excitation after extensive scattering, the excited-state number density ratio, $n_{2p}/n_{1s}$, approximately follows LTE conditions. This also applies to the $n_3/n_2$ ratio. Consequently, the $n_3/n_2$ ratio rises with temperature, boosting the H$\alpha$ source function and producing net emission in the line wings, which are formed at these depths.

\item \textit{NLTE effects suppress the source function.} 
Above $\sim$1.3 \Rp, the $n=3$ level population transitions from collision-dominated to radiation-dominated regimes as the number density becomes sufficiently low such that collisional excitation rates from $n=2$ to $n=3$ fall below the radiative excitation rates driven by stellar H$\alpha$.
The resulting NLTE departure drives $n_3/n_2$ downwards (Fig.~\ref{fig:best_fit}), while adiabatic expansion of the outflow further cools the gas. These effects create an absorbing layer above the emitting region.

\item \textit{Self-absorption carves the line core.} Because H$\alpha$ is optically thick, line-centre photons emitted from below are reabsorbed in the overlying layers; only photons with sufficient Doppler offsets escape. The line core probes $r\sim 1.8$ \Rp, while the wings probe $r\sim 1.3$ \Rp~for the $\dot{M}=10^{13}$ g\,s$^{-1}$ model (Fig.~\ref{fig:best_fit}). The contrast between emitting and absorbing layers produces the double-peaked morphology, and the specific line shape places stringent constraints on the vertical NLTE structure of the thermosphere.

\item \textit{Outflow winds create the peak asymmetry.} The vertical wind speed increases with altitude (Fig.~\ref{fig:best_fit}), so the absorbing layer moves preferentially towards the observer. This blueshifts the self-absorption relative to the underlying emission, suppressing the blue peak and producing the observed asymmetry. To reproduce the measured $\sim$0.5\% amplitude difference, the required outflow speed at the optical-depth-unity radius is $\sim$5 \kms.

\item \textit{Horizontal winds drive the phase variation.} In addition to vertical escape, \K~b has day-to-night winds reaching 12 \kms~in its lower atmosphere, as also inferred from the phase-dependent Doppler shifts of metal lines \citep{ZhangEtAl2026}. The H$\alpha$ emission shows a similar progressive redshift with phase (Fig.~\ref{fig:ha_main}b\&c). This shift arises as the planet rotates, an increasing fraction of the dayside hemisphere rotates into view, and the day-to-night winds advect atmospheric gas away from the observer (Fig.~\ref{fig:schematic}). The spectrally- and phase-resolved H$\alpha$ emission thus reflects a consistent picture of global atmospheric circulation with the metal lines.

\end{itemize}

Our 1D model reproduces the overall morphology and its phase variation, but the observed line is broader than the model prediction, hinting at underestimated thermal broadening, scattering of stellar H$\alpha$ radiation, or three-dimensional wind patterns that our one-dimensional treatment cannot capture. The plane-parallel radiative transfer also becomes less valid in the extended layers probed by the line core. These effects do not alter the double-peaked morphology, which is set by the vertical NLTE structure, but they do influence the detailed line shape. Future three-dimensional hydrodynamic simulations and NLTE radiative transfer calculations will be needed to fully exploit the constraining power of H$\alpha$ emission.

\cite{Sanchez-LopezEtAl2022} reported the transmission detection of Paschen $\beta$ line with a contrast of $\sim$0.53\%, a blueshift of $\sim$14.8 \kms. This line probes the same excited state $n=3$ as the H$\alpha$ emission line. \cite{Sanchez-LopezEtAl2022} interpreted the blueshift as the day-to-night flow in the bound atmosphere ($\sim$1.3 \Rp), which is consistent with our interpretation of the phase-dependent Doppler shifts observed in the emission line.
Alternatively, their study proposed that the origin of Pa$\beta$ absorption could be in a tail of escaping gas moving toward the observer. Although less probable, this scenario could potentially be viable if flaring events arising from star-planet interactions populate the H($n=3$) in the escaping tail \citep{Sanchez-LopezEtAl2022}.
Future modeling work combining all available observations of the absorption lines (Balmer and Paschen series) and the H-alpha emission will provide tight constraints on the vertical structure of the H population.

\section{Implications for mass loss} \label{sec:massloss}

The emission profile lifts the temperature--mass-loss-rate degeneracy that persists in transmission-based studies of atmospheric escape.
Previous H$\alpha$ transmission analyses in \K~b inferred $\dot{M}\sim10^{12}$--$10^{12.8}$ g\,s$^{-1}$ \citep{YanHenning2018, WyttenbachEtAl2020}; the recent Mg\,{\sc ii}/Fe\,{\sc ii} near-ultraviolet (NUV) measurement gave $\dot{M}\sim10^{12}$--$10^{12.5}$ g\,s$^{-1}$ \citep{BaldwinEtAl2025}; and self-consistent models predict up to $\sim 10^{12.5}$--$10^{13}$ g\,s$^{-1}$ for a 2 \Mjup~to 1.2 \Mjup~planet \citep{MunozSchneider2019}.
The planet mass itself is poorly known because of the star's rapid rotation, with literature values ranging from $2.17\pm0.56$ to $2.88\pm0.35$ \Mjup~\citep{GaudiEtAl2017, BorsaEtAl2019, HoeijmakersEtAl2019, PaiAsnodkarEtAl2022}, compounding the uncertainty.

Our emission-based analysis requires $\dot{M}\sim10^{13}$ g\,s$^{-1}$ or higher to produce the double-peaked line morphology (Fig.~\ref{fig:model_grid}): higher mass-loss rates (and the accompanying higher densities and pressures) push the LTE-to-NLTE transition outward, which is essential to produce the unique source-function profile in Fig.~\ref{fig:best_fit}.
Mass-loss rates much above $10^{13}$ g\,s$^{-1}$ would require an unphysically high base pressure ($P>10$ bar) at 1 \Rp, inconsistent with the expected photospheric level.
Our value therefore sits at the upper end of the literature range and favours a planet at the low end of the reported mass interval.
This mass-loss rate of $10^{13}$ g\,s$^{-1}$ means that the planet could lose $\sim$8\% of its total mass in 1 Gyr.

This is at the extreme high end of reported atmospheric escape rates for the broader population of transiting planets, such as warm Neptunes, hot Jupiters, and other ultrahot Jupiters ($\dot{M}\sim10^{8}$--$10^{12}$ g\,s$^{-1}$) \citep{EhrenreichEtAl2015, BourrierEtAl2018, EtangsEtAl2010, ZhangEtAl2023, CzeslaEtAl2024, AllartEtAl2025}.
Notably, the high mass-loss rates tend to be found in planets with high Roche filling factors ($R_\mathrm{p}/R_\mathrm{Hill}$) orbiting early-type stars, which facilitate atmospheric escapes \citep{SaidelEtAl2025}.
Compared to typical ultrahot Jupiters orbiting F-type stars, \K~b has a relatively lower Roche filling factor of $\sim$0.39 (versus $\sim$0.5 for WASP-121 b) and weaker X-ray and extreme ultraviolet (XUV) irradiation from its A-type host star.
Therefore, the higher mass-loss rate in \K~b supports the alternative Balmer-driven escape mechanism that the stellar NUV irradiation can power the outflow in planets orbiting A-type stars \citep{MunozSchneider2019}.
Future emission models accounting for 3D effects will allow for improved constraints on the mass-loss rate.

\section{Conclusion}

We present the detection of H$\alpha$ (6564.6 \AA) emission line from the dayside of the ultra-hot Jupiter \K~b using the high-resolution spectrometer KPF on the Keck I telescope.
This represents the first hydrogen emission detected in any escaping atmosphere.
Intriguingly, the spectrally resolved H$\alpha$ emission shows double peaks with amplitudes of $0.1-0.15\%$ at $\pm30$ \kms~along with an absorption feature in the line center. 
We apply NLTE outflow modeling using customized \texttt{sunbather} code and compute synthetic spectra to interpret the observed line profile.

This double-peaked line profile is a consequence of the extreme atmospheric condition, which leads to an increase of the excited-state H\,{\sc i} $n_3/n_2$ ratio due to strong thermal inversion in the lower atmosphere and a fall-off due to NLTE effect in the upper atmosphere.
This causes the self-absorption of the H$\alpha$ emission in the upper atmosphere, resulting in the double-peaked line shape.
We find that a large mass-loss rate of $\dot{M}\sim 10^{13}$ g\,s$^{-1}$ is required to produce the line profile. 
It also provides key dynamical information on the escaping atmosphere: the asymmetric double peaks indicate a vertical outflow wind speed of $\sim$5 \kms~at the altitude probed by the H$\alpha$ line core; the phase-dependent Doppler shift points to a horizontal day-to-night winds of $\sim$12 \kms.

The detection of spectrally- and phase-resolved hydrogen airglow from \K~b demonstrates that the thermal emission is a powerful tracer that can probe the atmospheric escape of short-period exoplanets.
The double-peaked line morphology directly records the vertical thermal structure, the NLTE departure of the excited-state populations, the outflow wind speed, and the day-to-night circulation---quantities that transmission alone cannot separate.
Extending this diagnostic to other hot Jupiters, and to other transitions accessible from the ground, will open a new observational window on the escape and evolution of worlds in extreme radiation environments.





\begin{acknowledgements}
    We thank Antonija Oklop\v{c}i\'c and Ignas Snellen for their helpful discussion.
    Y.Z. acknowledges the support from the Heising-Simons Foundation 51 Pegasi b Fellowship (grant \#2023-4298).
    A.H. acknowledges support from the National Science Foundation Graduate Research Fellowship under Grant No. 2141064 and the MIT Dean of Science Fellowship.
    This work used observations obtained at the W.M. Keck Observatory.
    The computations were carried out on the Caltech High-Performance Cluster.
\end{acknowledgements}

\vspace{5mm}
\facilities{KECK:I/KPF}

\software{
\texttt{numpy} \citep{HarrisEtAl2020},
\texttt{scipy} \citep{VirtanenEtAl2020},
\texttt{matplotlib} \citep{Hunter2007},
\texttt{astropy} \citep{astropy:2013, astropy:2018, astropy:2022},
\texttt{sunbather} \citep{LinssenEtAl2024},
\texttt{p-winds} \citep{DosSantosEtAl2022},
\texttt{Cloudy} \citep{GunasekeraEtAl2023},
\texttt{Molecfit} \citep{SmetteEtAl2015},
\texttt{easychem} \citep{LeiMolliere2024}.
}

\appendix
\restartappendixnumbering

\section{Supplementary plots}
\label{sec:sup_plots}

\begin{figure}[ht]
\centering
    \includegraphics[width=0.95\linewidth]{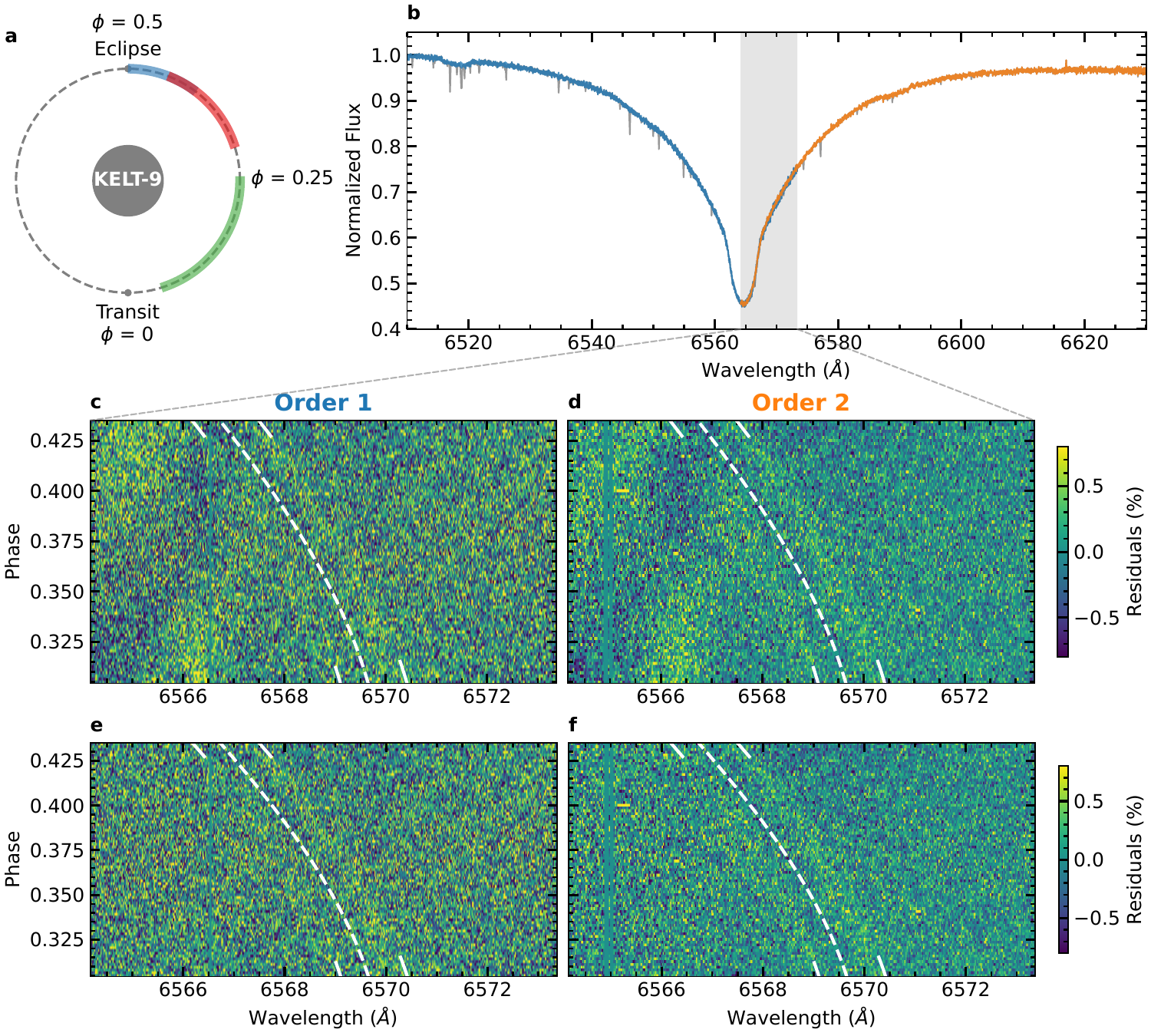}
    \caption{Phase coverage of KPF observations and time-series spectra of \K~b.
    \textbf{a}, Phase coverage of three epochs of KPF observations (shaded curves). The illustrations of object and orbit sizes are to scale.
    \textbf{b}, One exposure of the \K~spectrum around H$\alpha$ as observed with KPF. Blue and orange show two different spectral orders with an overlapping wavelength region (grey shading). Grey lines in the background show the spectrum before telluric removal.
    \textbf{c}, \textbf{d}, One epoch of 2D time-series residual spectra after removing the average stellar spectrum in each order. White dashed lines mark the expected centre of the planetary H$\alpha$ emission following the Keplerian orbital motion. The residual pattern at bluer wavelengths is caused by stellar pulsation contamination.
    \textbf{e}, \textbf{f}, Same as panel \textbf{c} and \textbf{d}, but removing the stellar contamination using Gaussian Processes for visual clarity. During the data analysis we masked the contaminated region with $|v|<110$ \kms~in the stellar rest frame.
    \label{fig:app1}}
\end{figure}

\begin{figure}[ht]
\centering
    \includegraphics[width=0.55\linewidth]{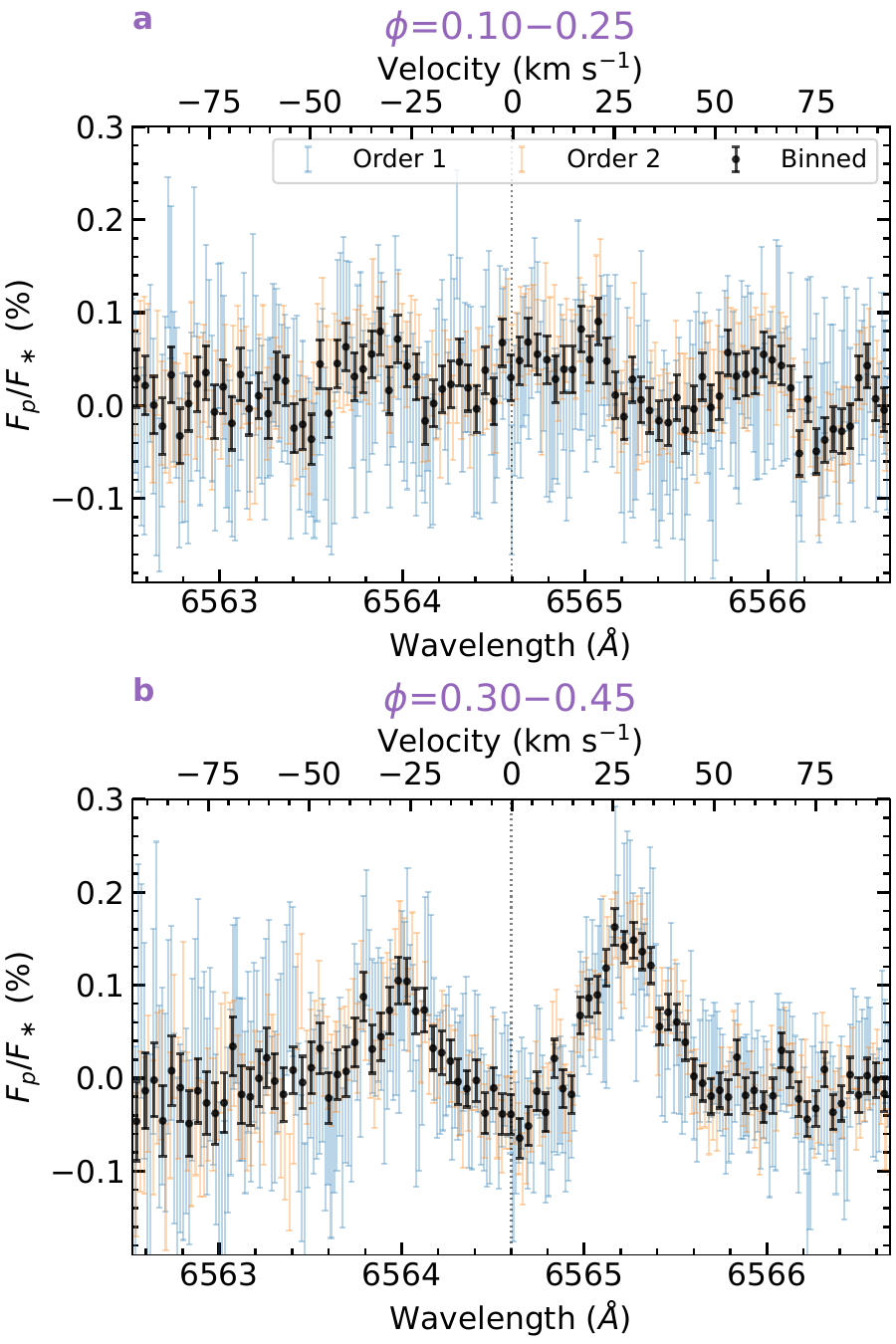}
    \caption{Phase-averaged H$\alpha$ line profile in the planet's rest frame.
    \textbf{a}, $\phi=0.05$--$0.25$, where the dayside is largely hidden and no significant signal is detected.
    \textbf{b}, $\phi=0.30$--$0.45$, where the dayside dominates and the double-peaked emission with central self-absorption is detected at $10\sigma$. Blue and orange points show the two independent spectral orders; black points are binned fluxes in every five data points.
    \label{fig:ha_1d}}
\end{figure}

\begin{figure}[ht]
\centering
    \includegraphics[width=0.6\linewidth]{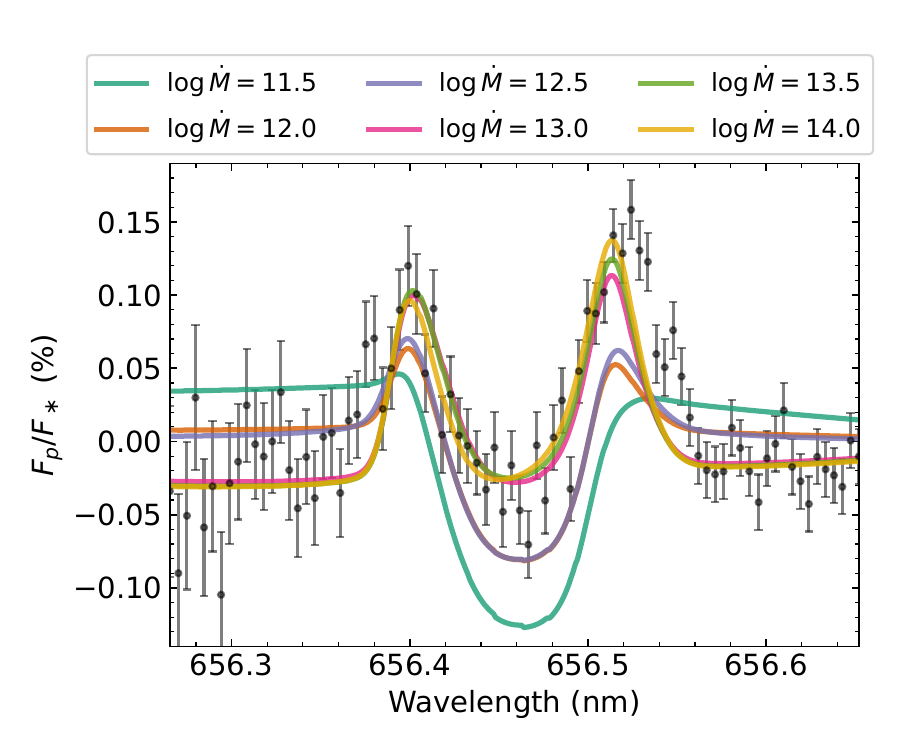}
    \caption{NLTE outflow model spectra as a function of mass-loss rate.
    Synthetic H$\alpha$ emission profiles for mass-loss rates $\dot{M}$ ranging over three orders of magnitude. Mass-loss rates above $10^{13}$ g\,s$^{-1}$ are required to produce the double-peaked morphology; higher values yield more asymmetric double peaks due to stronger vertical outflow winds.
    \label{fig:model_grid}}
\end{figure}

We present the phase coverage of our KPF observations in Fig.~\ref{fig:app1} panel a, an example of the observed spectrum and telluric removal in panel b, and the 2D time-series spectra in stellar rest frame in panel c, d, e, and f.
Fig.~\ref{fig:ha_1d} shows the phase-averaged H$\alpha$ emission detection in \K~b. 
The synthetic NLTE outflow model spectra for various mass-loss rates ranging from $10^{11.5}$ to $10^{14}$ g\,s$^{-1}$ are shown in Fig.~\ref{fig:model_grid}.

\bibliography{manuscript}
\bibliographystyle{aasjournal}

\end{document}